\documentclass[final,5p,times,twocolumn,numbers]{elsarticle} 

\usepackage{amssymb}
\usepackage{lipsum}
\usepackage{amsmath}
\usepackage{graphicx}
\usepackage{adjustbox}
\usepackage{subcaption}
\usepackage{caption}
\usepackage[colorlinks=true, allcolors=blue]{hyperref}
\usepackage{dcolumn}
\usepackage{bm}
\usepackage{color}

\biboptions{sort,compress} 
\journal{Physics Letters B}

\begin{document}

\begin{frontmatter}



\title{Observation of renormalization group invariance \\
in symmetry-restored nuclear lattice effective field theory}


\author[label1]{Jia-Ai Shi}
\affiliation[label1]{organization={Graduate School of China Academy of Engineering Physics},
            city={Beijing},
            postcode={100193}, 
            country={China}}

\author[label1]{Chen-Can Wang}

\author[label1]{Bing-Nan Lu\corref{cor1}}
\cortext[cor1]{bnlv@gscaep.ac.cn}

\begin{abstract}
Renormalization group (RG) invariance implies that the predictions of effective field theory are independent of the momentum cutoffs introduced during regularization.
Here we report the first systematic verification of RG invariance for realistic nuclear few-body systems within nuclear lattice effective field theory.
To restore broken continuum rotational and Galilean symmetries, we employ Galilean-invariance-restoration counterterms and use a soft momentum regulator. 
We calibrate the two- and three-body next-to-next-to leading order (N$^2$LO) chiral forces using $A\leq 3$ observables and perform precision quantum Monte Carlo calculations to compute the $^4$He binding energy.
The predicted energy remains constant across cutoffs from $250$~MeV to $400$~MeV and agrees well with the experimental value, with discrepancies of order 100 keV. 
Our results demonstrate the capability of extracting accurate, cutoff-independent predictions within lattice-regulated \textit{ab initio} nuclear theory.



\end{abstract}




\end{frontmatter}




\section{Introduction}
\label{introduction}


As a versatile \textit{ab initio} framework for nuclear physics, nuclear lattice effective field theory (NLEFT) has been applied to study diverse phenomena starting from fundamental inter-nucleon forces~\cite{lee2004nuclear, lee2005neutron, lee2009lattice}. In this approach, the nuclear quantum many-body problem is discretized on a cubic spatial lattice and solved using direct diagonalization or auxiliary field Monte Carlo methods. Recent NLEFT applications include studies of nuclear ground and excited states~\cite{borasoy2007lattice,epelbaum2010lattice1,epelbaum2010lattice,epelbaum2014ab,lahde2014lattice,lu2019essential,lu2022perturbative,shen2023emergent,elhatisari2024wavefunction, meissner2024ab,Shen:2024qzi}, nuclear clustering~\cite{epelbaum2011ab,epelbaum2012structure,epelbaum2013viability,elhatisari2017ab,Zhang:2024wfd}, scattering processes~\cite{bour2012benchmark,elhatisari2015ab}, finite-temperature nuclear matter~\cite{elhatisari2016nuclear,lu2020ab,ren2024ab,ma2024structure}, and hypernuclei~\cite{bour2015ab,scarduelli2020method,hildenbrand2024towards}. Overall, these calculations have shown excellent agreement with experimental data and provided deep insights into the mechanisms underlying complex strong-correlation phenomena.

Despite these achievements, most NLEFT calculations employ a fixed lattice spacing, typically between 1 fm and 2 fm (corresponding to momentum cutoffs of 314 MeV to 628 MeV). 
According to effective field theory (EFT) principles, it is crucial to systematically verify that NLEFT predictions are independent of the lattice spacing (that is, to establish their renormalization group (RG) invariance). 
However, such calculations at multiple resolutions remain computationally expensive and have so far been largely confined to two-body systems. 
Examples include the construction of lattice nucleon-nucleon interactions with systematic spacing variations~\cite{alarcon2017neutron, li2018neutron, wu2025charge} and studies of their renormalization in the two-body sector~\cite{KLEIN2015511, Korber:2019cuq, chandrasekharan2024worldline}. 
Beyond two-body systems, systematic investigations have been limited primarily to few-body calculations of the Tjon line~\cite{klein2018tjon}. 
These studies revealed that the empirical linear correlation between $^3$H and $^4$He binding energies diminishes at larger lattice spacings but is recovered at smaller ones. 
However, the $^4$He results in these calculations used a first-order perturbative approximation and did not fully account for symmetry-breaking effects. 
This lack of precision limits the ability to definitively establish RG invariance. 
In contrast, studies of one-dimensional bosons with zero-range contact interactions employed improved lattice actions, demonstrating systematic order-by-order improvement in the EFT expansion towards the continuum limit~\cite{EPJA54-233}. 
Corresponding investigations using realistic chiral forces face significant challenges and are still lacking. 
The primary challenges include the high computational cost of high-precision calculations and the complicated operator structures involving intricate spin and isospin dependencies.

The renormalization of nuclear EFT in the continuum has been a long-standing issue since its initial development for constructing nuclear forces~\cite{weinberg1991effective, weinberg1992three}.
In the EFT framework, several momentum cutoffs separate physical low-momentum degrees of freedom from irrelevant high-momentum modes.
Predictions from EFT should be independent of the specific cutoff value within controlled systematic uncertainties~\cite{epelbaum2009modern, machleidt2011chiral,RevModPhys.92.025004, epelbaum2020high}.
According to Wilsonian RG theory, irrelevant degrees of freedom are integrated out into cutoff-dependent parameters called low-energy constants (LECs)~\cite{kvinikhidze2007wilsonian,harada2013wilsonian, doi:10.1142/S021830131641007X, epelbaum2017wilsonian, Epelbaum_2018, kvinikhidze2018renormalisation}.
RG invariance requires readjusting the LECs to compensate for cutoff variations. 
Current RG analyses in nuclear EFT have predominantly focused on two- and three-body systems~\cite{kaplan1996nucleon,kaplan1998new,  bedaque1999renormalization, bedaque1999three, BEDAQUE2000357, epelbaum2017renormalization, epelbaum2018not,epelbaum2020renormalize, gasparyan2023renormalization}.
Investigations extending to heavier nuclei like $^4$He and $^{16}$O have been recently persued using modern many-body approaches~\cite{PhysRevA.70.052101,PLATTER2005254,refId0, KIRSCHER2013335, PhysRevLett.122.143001,  yang2023importance, Contessi:2025xue}.
Despite this, RG invariance of nuclear EFT in many-body systems remains inadequately established. 
Note that the lattice formulation of EFT can yield outcomes substantially different from its continuum counterpart, as demonstrated in recent calculations of the $^4$He monopole transition form factor~\cite{PhysRevLett.110.042503, PhysRevLett.130.152502, meissner2024ab}. It is therefore valuable to conduct separate and in-depth studies of the distinct properties of the lattice formalism, particularly its renormalization. A clear understanding of EFT renormalization on the lattice would not only improve the precision of NLEFT calculations but also offer an alternative perspective for addressing the challenges in renormalizing continuum EFTs.

In this work, we focus on the lattice formalism and systematically investigate the dependence of the few-body observables ($A \leq 4$) on the cutoff.
One of the essential differences between the lattice and continuum regulators is that the cubic lattice violates several critical symmetries such as the rotational symmetry and Galilean invariance.
These lattice artifacts introduce unphysical contamination into NLEFT predictions and should be eliminated before comparing with the experiments.
This has been achieved through either numerical averaging over symmetry groups like SO(3)~\cite{lu2014breaking, lu2015breaking} or supplementing the Hamiltonian with specific counterterms to restore symmetries~\cite{EPJA54-233, li2019galilean, elhatisari2024wavefunction}. 
The latter approach resembles the Symanzik improvement scheme widely used in lattice quantum chromodynamics (QCD), which introduces irrelevant operators to cancel lattice artifacts and accelerate convergence toward the continuum limit~\cite{symanzik1983continuum, symanzik1983continuum2}. 
Although in NLEFT it remains far from clear whether there is a well-defined continuum limit, we can nevertheless investigate how symmetry breaking and restoration affect the evolution of the observables.

Here we employ nuclear chiral forces up to next-to-next-to-leading order (N$^2$LO), which provide sufficient accuracy for light nuclei with $A\leq 4$. 
Compared to Ref.~\cite{klein2018tjon}, we implement two significant improvements. 
First, we utilize a soft lattice regulator to mitigate rotational symmetry breaking effects, thereby isolating the consequences of Galilean invariance breaking~\cite{KLEIN2015511, EPJA54-233}. 
This simplifies the analysis, and the ensuing conclusions are directly applicable to ordinary lattice regulators. 
Second, we apply the recently developed perturbative quantum Monte Carlo (ptQMC) method~\cite{lu2022perturbative, liu2025perturbative}.
This method circumvents the Monte Carlo sign problem, enabling the calculation of $^4$He binding energies up to second-order perturbation theory.
This approach delivers substantially higher precision than the first-order perturbative calculations employed in Ref.~\cite{klein2018tjon}, typically achieving 
$~$0.1~MeV accuracy for $^4$He, facilitating quantitative investigation of subtle lattice effects.

The paper is organized as follows. 
Section II details the lattice interaction and regularization scheme. 
Section III presents low-energy constant (LEC) determination and compares NLEFT predictions with and without symmetry restoration. 
We conclude with a summary and discussion of our results' implications.

\section{Theoretical framework}
We employ the next-to-next-to-leading-order (N$^2$LO) lattice chiral force introduced in Ref.~\cite{lu2022perturbative}. 
To minimize lattice artifacts, we implement all spatial derivatives via fast Fourier transform (FFT) instead of finite-difference schemes, 
and introduce an isotropic low-momentum cutoff to restore the rotational symmetry. 
Details of this interaction are provided in Ref.~\cite{liu2025perturbative}, but we summarize its momentum-space form here for completeness. 
We define the incoming and outgoing momenta as $\bm{p}_{1,2}$ and $\bm{p}^\prime_{1,2}$, respectively. 
The relative momenta are $\bm{p}=(\bm{p}_1 - \bm{p}_2)/2$ and $\bm{p}^\prime=(\bm{p}^\prime_1 - \bm{p}^\prime_2)/2$, 
the momentum transfers are $\bm{q}=\bm{p}^\prime - \bm{p}$ 
and $\bm{k} = (\bm{p}^\prime + \bm{p})/2$.
The spin and isospin Pauli matrices are denoted by $\boldsymbol{\sigma}_{1,2}$ and $\boldsymbol{\tau}_{1,2}$, respectively.

The leading order ($Q^0$) and next-to-leading order ($Q^2$) two-body contact terms are:
\begin{eqnarray}
V_{Q^0} &=& \bigl[ B_1 + B_2 (\boldsymbol{\sigma}_1 \cdot \boldsymbol{\sigma}_2) \bigr] f_{\rm 2N}(\{p_i, p_i^\prime\}),  \\\nonumber
V_{Q2} &=& \bigl[ C_1 q^2 + C_2 q^2 (\boldsymbol{\tau}_1 \cdot \boldsymbol{\tau}_2) +C_3 q^2 (\boldsymbol{\sigma}_1 \cdot \boldsymbol{\sigma}_2) \bigr.  \\\nonumber
 & & + C_4 q^2 (\boldsymbol{\sigma}_1 \cdot \boldsymbol{\sigma}_2) (\boldsymbol{\tau}_1 \cdot \boldsymbol{\tau}_2)  \\\nonumber
 & & + C_5\frac{i}{2} (\boldsymbol{q} \times \boldsymbol{k}) \cdot (\boldsymbol{\sigma}_1 + \boldsymbol{\sigma}_2) 
  + C_6 (\boldsymbol{\sigma}_1 \cdot \boldsymbol{q}) (\boldsymbol{\sigma}_2 \cdot \boldsymbol{q}) \\\nonumber
  & & \bigl. + C_7 (\boldsymbol{\sigma}_1 \cdot \boldsymbol{q}) (\boldsymbol{\sigma}_2 \cdot \boldsymbol{q}) (\boldsymbol{\tau}_1 \cdot \boldsymbol{\tau}_2) \bigr] f_{\rm 2N}(\{p_i, p_i^\prime\}),  
\label{eq:N2LOcontacts}
\end{eqnarray}
where \( B_{1-2} \) and \( C_{1-7} \)  are LECs.
For all two-body contact terms, we have applied a multiplicative single-particle regulator 
\begin{eqnarray}
\begin{aligned}
 f_{\rm 2N}(\{p_i, p_i^\prime\}) = \exp \left[ - \sum_{i=1}^2 \left( p_i^6 + p_i^{\prime 6} \right) / (2 \Lambda^6) \right].
 \label{eq:lat_reg}
\end{aligned}
\end{eqnarray}
which is implemented by transforming the single-particle wave functions to momentum space using FFT and multiplying them by the regulator function Eq.~(\ref{eq:lat_reg}).

For long-range interactions, we include the local one-pion-exchange potential (OPEP)~\cite{reinert2018semilocal},
\begin{eqnarray}
\begin{aligned}
V_{1\pi} =-\frac{g_A^2 f_\pi \left( q^2 \right)}{4F_\pi^2} \left[ \frac{(\boldsymbol{\sigma}_1 \cdot \boldsymbol{q}) (\boldsymbol{\sigma}_2 \cdot \boldsymbol{q})}{q^2 + M_\pi^2} + C_\pi' \boldsymbol{\sigma}_1 \cdot \boldsymbol{\sigma}_2 \right] (\boldsymbol{\tau}_1 \cdot \boldsymbol{\tau}_2), \nonumber
\end{aligned}
\end{eqnarray}
where $g_A = 1.287$, $F_\pi = 92.2$~MeV, and $M_\pi = 134.98$~MeV are the axial coupling constant, pion decay constant, and pion mass, respectively. 
The OPEP regulator employs an exponential form,
$f_\pi(q^2) = \exp\left[ - \left( q^2 + M_\pi^2 \right) / \Lambda_\pi^2 \right]$, 
with $\Lambda_\pi$ denoting the pion momentum cutoff. 
The constant \( C_\pi' \) is defined as
\begin{eqnarray}
 \begin{aligned}
 C_\pi' = -\left[ \frac{\Lambda_\pi^2 - 2 M_\pi^2}{3 \Lambda_\pi^2} + 2 \sqrt{\pi} \frac{M_\pi^3}{3 \Lambda_\pi^3} \exp \left( \frac{M_\pi^2}{\Lambda_\pi^2} \right) \text{erfc} \left( \frac{M_\pi}{\Lambda_\pi} \right) \right]. \nonumber
 \end{aligned}
\end{eqnarray}
The term proportional to \( C_\pi' \) is introduced to eliminate the short-range singularity. 
The OPEP plays an essential role in reproducing the NN phase shifts below $200$~MeV~\cite{alarcon2017neutron, li2018neutron, wu2025charge}.

For protons, we include a regularized Coulomb force
\begin{equation}
V_{\rm cou} = \frac{\alpha}{q^2}  \left(\frac{1 + \tau_{1z}}{2}\right) \left(\frac{1 + \tau_{2z}}{2}\right)  f_{\rm cou}(q^2),
\label{eq:CoulombPotential}
\end{equation}
where $\alpha=1/137$ is the fine structure constant, $f_{\rm cou}(q^2) = \exp\left(-q^2 / \Lambda_{\rm cou}^2\right)$ is the Coulomb regulator, with $\Lambda_{\rm cou}$ the photon momentum cutoff.
Residual dependencies on $\Lambda_{\rm cou}$ are absorbed into the leading-order charge-symmetry-breaking (CSB) contact terms,
\begin{align}
V_{{\rm CSB}}^{{\rm pp}} & =c_{{\rm pp}}\left(\frac{1+\tau_{1z}}{2}\right)\left(\frac{1+\tau_{2z}}{2}\right)f_{\rm 2N}(\{p_i, p_i^\prime\}),\nonumber \\
V_{{\rm CSB}}^{{\rm nn}} & =c_{{\rm nn}}\left(\frac{1-\tau_{1z}}{2}\right)\left(\frac{1-\tau_{2z}}{2}\right)f_{\rm 2N}(\{p_i, p_i^\prime\}),\label{eq:VCSB}
\end{align}
where $c_{\rm pp, nn}$ are LECs and we have applied the same regulator Eq.~(\ref{eq:lat_reg}) as for the normal contact terms.

For the three-body force, we adopt a simple three-body contact term appearing at N$^2$LO,
\begin{align}
 V_{\rm 3N} = \frac{c_E}{2F_\pi^4 \Lambda_\chi} f_{\rm 3N} (\{p_i, p_i^\prime\} ),    
\end{align}
where \( c_E \) is a LEC, $\Lambda_\chi = 700$~MeV represents the chiral symmetry breaking scale, 
\begin{eqnarray}
\begin{aligned}
 f_{\rm 3N}(\{p_i, p_i^\prime\}) = \exp \left[ - \sum_{i=1}^3 \left( p_i^6 + p_i^{\prime 6} \right) / (2 \Lambda^6) \right] , 
 \label{eq:lat_reg3}
\end{aligned}
\end{eqnarray}
is a separable single-particle regulator.
In this work we always use the same value of $\Lambda$ in both two- and three-body regulators,
enabling the simultaneous auxiliary-field transformation of both two- and three-body forces.

The regulators Eq.~(\ref{eq:lat_reg}) and Eq.~(\ref{eq:lat_reg3}) act on single-particle momenta rather than relative momenta $\bm{p}$ and $\bm{p}^\prime$, thus breaking Galilean invariance.
Nevertheless, for NN scattering, we typically restrict the analysis to the center-of-mass frame, where 
\begin{align}
  \boldsymbol{p} = \boldsymbol{p}_1 = -\boldsymbol{p}_2 \quad \mathrm{and} \quad \boldsymbol{p}' = \boldsymbol{p}_1' = -\boldsymbol{p}_2'.
\end{align}
In this case Eq.~(\ref{eq:lat_reg}) is equivalent to the non-local Galilean-invariant regulator commonly employed in chiral force constructions in the continuum~\cite{entem2003accurate,epelbaum2005two,  EPJA51-53},
\begin{equation}
  f_{\rm 2N}^{\rm rel} = \exp \left( - \left( p^6 + p'^6 \right) / \Lambda^6 \right).
  \label{eq:continuum_reg}
\end{equation}
Consequently, when the interactions are parametrized in the center-of-mass frame, the resulting two-body LECs are identical irrespective of the regulator used.
However, for calculations in moving frames or in many-body systems ($A\geq 3$), the interacting nucleon pair may possess a nonzero total momentum. 
Under these conditions, the two regulators are no longer equivalent and generally yield different results.

Differently from the case of the rotational symmetry, so far there is no Galilean-invariant implementation of the contact terms on the lattice.
Here we follow Ref.~\cite{li2019galilean,elhatisari2024wavefunction} to restore Galilean invariance by introducing a set of Galilean-invariance-restoration (GIR) counterterms to the Hamiltonian.
Here we implement them directly in momentum space via FFT,
\begin{eqnarray}
V_{\rm GIR} = \left[ g_1 Q^2 + g_2 Q^2 (\bm{\sigma}_1 \cdot \bm{\sigma}_2)\right] f_{\rm 2N}(\{p_i, p_i^\prime\}),
\label{eq:GIRterms}
\end{eqnarray}
where $g_{1,2}$ are LECs, and $\bm{Q} = \bm{p}_1 + \bm{p}_2 = \bm{p}_1^\prime + \bm{p}_2^\prime$ denotes the pair's total momentum.
Here we employ the same single-particle regulator Eq.~(\ref{eq:lat_reg}) and cutoff value $\Lambda$ as those used for the two- and three-body contact terms.

In summary, the total Hamiltonian is given by
\begin{eqnarray}
H &=& K + V_{\rm Q^0} + V_{\rm Q^2} + V_{1\pi} + V_{\rm cou} \\\nonumber
& & + V_{\rm CSB}^{\rm pp} + V_{\rm CSB}^{\rm nn} + V_{\rm GIR} + V_{\rm 3N},
\end{eqnarray}
where $K$ denotes the kinetic energy term.
The contact terms $V_{\rm Q^0}$, $V_{\rm Q^2}$, $V_{\rm CSB}^{\rm pp, nn}$, $V_{\rm GIR}$, and $V_{\rm 3N}$ are all regulated based on single-particle momenta with the same cutoff $\Lambda$. 
In contrast, the long-range terms $V_{1\pi}$ and $V_{\rm cou}$ are regulated based on the momentum transfer $\bm{q}$, with cutoffs $\Lambda_\pi$ and $\Lambda_{\rm cou}$, respectively.
This study focuses on the renormalization of the short-range lattice interactions, leaving detailed discussions for lattice-regulated pion-exchange and Coulomb potentials to future work.

\section{Results and discussion}


\subsection{Determination of LECs}
\label{Sec:3A}

Throughout this work, all calculations employ a fixed lattice spacing $a=0.987$~fm, corresponding to an anisotropic momentum cutoff $\Lambda_a \sim \pi/a \approx 628$~MeV. 
The cutoffs for OPEP and Coulomb potential are fixed to $\Lambda_{1\pi} = \Lambda_{\rm cou} = 300$~MeV.
The contact term cutoff $\Lambda$ is systematically varied from 250~MeV to 400~MeV with 25~MeV increments.
We choose such a cutoff range adequate to demonstrate the RG invariance for light nuclei.
In similar studies, pionless EFT often varies the cutoff over a wide range across a few orders~\cite{PhysRevLett.122.143001, Contessi:2025xue}, whereas pion-full EFT studies are mostly confined to $\Lambda \lesssim 500$~MeV due to numerical challenges~\cite{yang2023importance}.
Given that $\Lambda_a$ substantially exceeds the other cutoffs, the low-energy physics is independent of the lattice spacing and solely governed by the smooth cutoffs $\Lambda_{1\pi}$, $\Lambda_{\rm cou}$, and $\Lambda$.
Here the lattice is invisible to the nucleons and we can safely take the continuum limit $a\rightarrow 0$.
Alternatively, we can also investigate the cutoff-dependence by removing the $\Lambda$-cutoff and directly varying the lattice spacing $a$.
However, this approach introduces significant rotational-symmetry breaking effects and discontinuous discretization errors, which substantially complicate the analysis.
The soft regulators introduced in Eq.~(\ref{eq:lat_reg}) and (\ref{eq:lat_reg3}) effectively suppress these inessential lattice artifacts, creating an optimized framework for isolating and analyzing Galilean invariance violations~\cite{KLEIN2015511, EPJA54-233}.
The effects of rotational symmetry breaking can be investigated afterwards once Galilean invariance restoration is fully characterized.

In this exploratory study, we investigate light nuclei $^3$H, $^3$He, and $^4$He with $A \leq 4$.
For $^3$H and $^3$He, we diagonalize the lattice Hamiltonian using the Lanczos method, while for the $^4$He nucleus, we employ the recently developed perturbative quantum Monte Carlo (ptQMC) approach~\cite{lu2022perturbative}.
Within the ptQMC framework, nuclear binding energies are expanded around non-perturbative results from a Wigner-SU(4) action~\cite{lu2019essential} up to second order, effectively circumventing the fermionic sign problem inherent in quantum Monte Carlo simulations.
This method achieves high-precision solutions of chiral forces up to N$^3$LO for deeply bound systems such as $^4$He~\cite{liu2025perturbative}.
In this work, all calculations employ periodic boundary conditions.
To eliminate the finite volume effects, for $A=3$ systems we take $L=11,12,...15$ and extrapolate to the infinite volume limit, while for $^4$He we typically found convergence at $L=12$.

For NLEFT calculations, the LECs must first be determined.
For each $\Lambda$ value, we calibrate the LECs in $V_{\rm Q^0}$ and $V_{\rm Q^2}$ to reproduce empirical neutron-proton phase shifts in the center-of-mass frame~\cite{stoks1993partial}.
Phase shift calculations follow Ref.~\cite{lu2016precise}, utilizing lattice scattering wavefunctions decomposed into partial waves with auxiliary potentials extracting asymptotic radial wavefunctions.
Partial-wave phase shifts and mixing angles are determined precisely by comparing scattered and free wave solutions.

The spectroscopic LECs for each partial wave are optimized through $\chi^2$-minimization against the Nijmegen phase shift analysis using the Levenberg-Marquardt algorithm.
Subsequently, the standard LECs $B_{1-2}$ and $C_{1-7}$ are obtained via linear recombination of these spectroscopic LECs, following the transformation derived for continuum contact terms~\cite{epelbaum2005two}.
This calibration methodology has been successfully implemented in constructing high-fidelity N$^2$LO~\cite{alarcon2017neutron} and N$^3$LO~\cite{li2018neutron, wu2025charge} lattice chiral interactions, 
where the RG-invariance or the cutoff-dependence in the two-body sector has been examined carefully.
We fit neutron-proton phase shifts up to relative momenta $P_{\rm rel} \leq 200$~MeV, which is sufficient for analyzing light nuclear structure with $A \leq 4$.
Within this low-momentum regime, the two-pion-exchange potential (TPEP) becomes indistinguishable from short-range contact terms and is therefore omitted, which has been numerically confirmed in studies with the TPEP explicitly incorporated~\cite{wu2025charge}.


Table~\ref{tab:LECs} presents the fitted LECs for different $\Lambda$ values, expressed in lattice units ($\hbar=c=a=1$). 
Empirical neutron-proton phase shifts are well reproduced for the interval of $\Lambda$ considered here.
The CSB coefficients $c_{\rm nn}$ and $c_{\rm pp}$ are determined by fitting to the experimental neutron-neutron (nn) and proton-proton (pp) scattering lengths, respectively. 
For pp scattering, we include the Coulomb potential according to Eq.~(\ref{eq:CoulombPotential}). 
Details of the fitting procedure follow Ref.~\cite{wu2025charge}. 

\begin{figure}[htbp]
\centering
    \includegraphics[width=1\columnwidth]{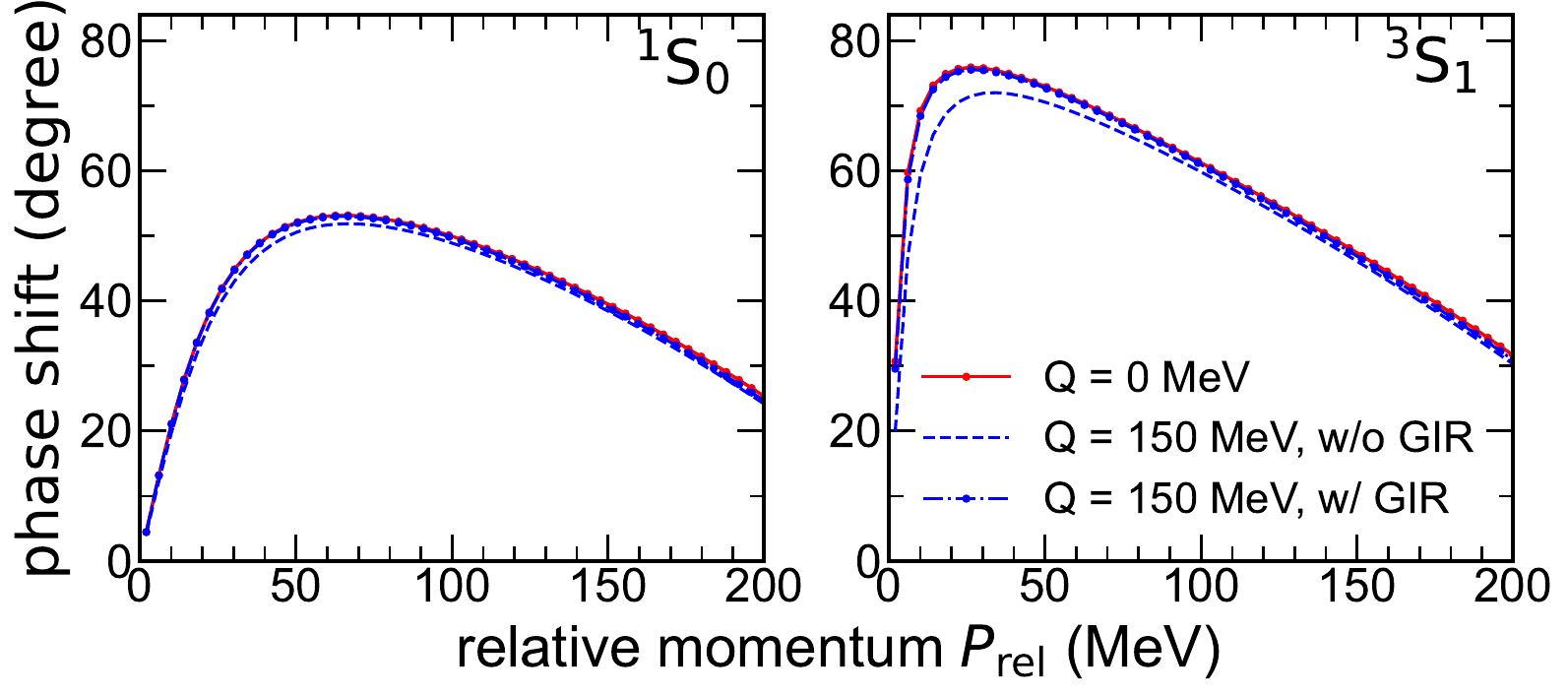}
\captionsetup{justification=raggedright, singlelinecheck=false}
\caption{\label{fig:GIR1} Phase shifts before and after adding GIR terms in $^1S_0$ state (left panel) and $^3S_1$ state (right panel).
$Q=0$ denotes the center of mass frame, where the GIR terms vanish completely.}
\end{figure}

The leading-order GIR coefficients $g_1$ and $g_2$ are determined by enforcing independence of the S-wave neutron-proton scattering lengths on the total momentum $\bm{Q}$. 
This approach follows Ref.~\cite{li2019galilean}. 
For S-waves, we average over total momentum orientations to obtain an effective $Q^2$-dependent potential. 
This dependence is eliminated by incorporating explicit $Q^2$-dependent contact terms as in Eq.~(\ref{eq:GIRterms}), which provide additional attraction in moving frames to compensate for the weakening of lattice-regulated contact terms.
Fig.~\ref{fig:GIR1} compare S-wave phase shifts up to $P_{\rm rel} \leq 200$~MeV at $Q=0$ and $Q=150$~MeV. 
While GIR terms have no effect in the stationary frame, they enable reproducing identical phase shift in moving frames.
The fitted $g_1$ and $g_2$ values against $\Lambda$ are given in Table~\ref{tab:LECs}. 
The results demonstrate a linear dependence of $g_1$ and $g_2$ on $\Lambda^{-2}$, confirming that Galilean invariance breaking effects scale as $\mathcal{O}(\Lambda^{-2})$ and diminish asymptotically with increasing $\Lambda$. 

Three-body forces have long been recognized as essential for reproducing observables across light nuclei to nuclear matter systems~\cite{RMP85-197}.
Conversely, interactions involving more nucleons such as four-body forces are generally considered negligible.
While this hierarchy aligns with effective field theory (EFT) power counting, quantitative understanding remains limited.
For simplicity, we restrict our analysis to a leading-order three-body contact term proportional to the dimensionless LEC $c_E$.
For each $\Lambda$ value, we first determine the two-body LECs using the aforementioned procedure, then calibrate $c_E$ to reproduce the experimental triton binding energy $E(^3\mathrm{H}) = -8.482$~MeV.
The resulting $c_E$ values are listed in Table~\ref{tab:LECs}.
To facilitate subsequent analysis, we additionally compute a parallel set of $c_E$ values excluding the two-body GIR terms, denoted as $c_E^\prime$ in Table~\ref{tab:LECs}.
As the contributions of the GIR term and three-body force are both attractive, generally we need stronger three-body force $c_E^\prime > c_E$ to compensate for the ommision of the GIR term.
Throughout this work, primed symbols always indicate calculations without Galilean invariance restoration.

\subsection{Predictions with sliding cutoff}
Next, we analyze the results calculated using the LECs fitted according to the previous section. We repeat calculations for different cutoffs $\Lambda$ and examine the dependencies of various observables on $\Lambda$. The target quantities used to calibrate the interactions remain constant, while the LECs and predicted observables are running functions of $\Lambda$. Quantifying these cutoff dependencies numerically is a primary focus of this work.

\begin{table*}
\renewcommand{\arraystretch}{1.3}
\captionsetup{justification=raggedright, singlelinecheck=false}
\caption{\label{tab:LECs} 
Fitted low-energy constants (LECs) and predicted observables for light nuclei systems, 
calculated with sliding momentum cutoff $\Lambda$. 
LECs $B_1$ through $c_E^\prime$ are given in natural lattice units ($\hbar=c=a=1$). 
Scattering length $a_0^\prime$ and effective range $r_0^\prime$ are calculated in the center-of-mass frame. 
Results labeled \texttt{2NF} are calculated using only two-nucleon forces, 
\texttt{2NF+3NF} include the three-nucleon force. 
Results marked with a prime (\textit{e.g.}, $c_E^\prime$) denote calculations without the Galilean-invariance-restoration (GIR) terms. 
$\Delta$ indicates the energy splitting between $^3\mathrm{H}$ and $^3\mathrm{He}$. 
All energies are in MeV, all lengths are in fm.
}
\begin{tabular}{ccccccccc}
 \textbf{$\Lambda$ (MeV)}  & \textbf{250} &\textbf{275} & \textbf{300}&\textbf{325}&\textbf{350}&\textbf{375}&\textbf{400}&\textbf{EXP}\\
\hline
    {$B_\mathrm{1}$}   &-4.931&-4.762&-4.592&-4.435&-4.285&-4.140&-3.997\\
    {$B_\mathrm{2}$}   &-0.369&-0.326&-0.285&-0.246&-0.206&-0.164&-0.119\\
    {$C_\mathrm{1}$}    &0.363&0.432&0.454&0.456&0.449&0.440&0.432\\
    {$C_\mathrm{2}$}    &0.063&0.011&-0.017&-0.032&-0.041&-0.048&-0.056\\
    {$C_\mathrm{3}$}   &-0.002&-0.024&-0.033&-0.039&-0.042&-0.047&-0.051\\
    {$C_\mathrm{4}$}    &0.008&-0.029&-0.049&-0.060&-0.067&-0.073&-0.078\\
    {$C_\mathrm{5}$}   
    &0.997&0.939&0.901&0.875&0.856& 0.841&0.829\\
    {$C_\mathrm{6}$}    & 0.024&0.015&0.007&0.002&-0.002&-0.004&-0.005\\
    {$C_\mathrm{7}$}   &-0.288&-0.267&-0.255&-0.248&-0.245&-0.245&-0.247\\

    {$c_\mathrm{nn}$}   &0.074    &0.068  &0.063 &0.060 &0.058 &0.057 &0.057 \\
    {$c_\mathrm{pp}$}   &0.174    &0.155  &0.142 &0.133 &0.127 &0.124 &0.124 \\
    
    {$g_1$}&-1.299   &-0.802 &-0.501&-0.312&-0.193&-0.119&-0.078\\
    {$g_2$}&-0.224   &-0.137 &-0.082&-0.048&-0.027&-0.014&-0.005\\  

    {$c_E$}   &1.863    &0.941  &0.504 &0.313 &0.245 &0.247 &0.289 \\
    {$c_E^\prime$}    &5.389    &2.928  &1.661 &0.995 &0.653 &0.496 &0.459 \\

\hline

        {$a_\mathrm{0}^\prime({^1S_0})$}   &-23.485&-23.517&-23.519&-23.539&-23.561&-23.567&-23.571&-23.74(2)\\    
    {$r_\mathrm{0}^\prime({^1S_0})$}   &2.424&2.452&2.449&2.471&2.497&2.504&2.507&2.77(5)\\
    {$a_\mathrm{0}^\prime({^3S_1})$}   &5.520&5.499&5.487&5.481&5.477&5.474&5.471&5.419(7)\\
    {$r_\mathrm{0}^\prime({^3S_1})$}   &1.588&1.616&1.631&1.643&1.650&1.655&1.658&1.753(8)\\   

    {$E^\prime_\mathrm{2NF}$($^3$H)} &-6.285  &-6.711&-7.096&-7.416&-7.642&-7.763&-7.774&-8.482\\
    {$E^\prime_\mathrm{2NF+3NF}$($^3$H)} &-8.482&-8.482&-8.482&-8.482&-8.482&-8.482&-8.482&-8.482\\
    {$E^\prime_\mathrm{2NF}$($^4$He)} &-19.21(12)&-20.30(3)&-21.63(3)&-22.89(6)&-23.75(8)&-24.24(7)&-24.24(6)&-28.30\\    {$E^\prime_\mathrm{2NF+3NF}$($^4$He)}&-30.78(16)&-30.13(12)&-29.89(15)&-29.65(13)&-29.36(13)&-29.25(13)&-29.07(13)&-28.30\\

\hline

    {$E_\mathrm{2NF}$($^3$H)}    &-7.542  &-7.789&-7.975&-8.080&-8.104&-8.054&-7.955&-8.482\\  
    {$E_\mathrm{2NF+3NF}$($^3$H)} &-8.482&-8.482&-8.482&-8.482&-8.482&-8.482&-8.482&-8.482\\ 
    {$E_\mathrm{2NF}$($^4$He)} &-24.13(5)&-24.81(5)&-25.54(7)&-26.05(9)&-26.15(9)&-25.87(8)&-25.37(7)&-28.30\\{$E_\mathrm{2NF+3NF}$($^4$He)}&-28.52(11)&-28.25(8)&-28.22(15)&-28.30(14)&-28.37(5)&-28.33(9)&-28.36(15)&-28.30\\
    {$E_\mathrm{2NF+3NF}$($^3$He)}  &-7.753 &-7.737&-7.723&-7.716&-7.710&-7.707&-7.703&-7.718\\
    {$\bigtriangleup $($^3$H - $^3$He)} &0.729   &0.745 &0.759&0.766&0.772&0.775&0.779&0.764\\ 
    {$R_\mathrm{2NF+3NF}$($^3$H)}&1.788(7)&1.758(6)&1.740(5)&1.729(7)&1.725(5)&1.721(4)&1.718(4)&1.759(36)\\
    {$R_\mathrm{2NF+3NF}$($^3$He)}&1.970(9)&1.960(2)&1.943(2)&1.935(9)&1.929(3)&1.936(5)&1.934(11)&1.966(3)\\

    {$R_\mathrm{2NF+3NF}$($^4$He)}&1.881(4)&1.777(4)&1.715(3)&1.684(3)&1.673(3)&1.674(4)&1.676(4)&1.676(3)\\


\hline

\end{tabular}
\end{table*}

The second part of Table~\ref{tab:LECs} (below $c_E^{\prime}$) presents various observables calculated without the GIR corrections. 
Here, $a_0^{\prime}$s and $r_0^{\prime}$s denote the $S$-wave scattering lengths and effective ranges in the center-of-mass frame, respectively. These results demonstrate near cutoff independence and are close to the experimental values. Deviations from experiments gradually decrease as $\Lambda$ increases, which can be further reduced by including next-to-next-to-next-to-leading order (N$^3$LO) contact terms and two-pion-exchange potentials~\cite{wu2025charge}.

For $^{3}$H and $^{4}$He nuclei, we present binding energies calculated without ($E_{\rm 2NF}^{\prime}$) and with ($E_{\rm 2NF+3NF}^{\prime}$) the three-body force, respectively. 
Without the three-body force, both nuclei are underbound, and their binding energies vary by approximately 20\% over the $\Lambda$ interval considered here. For the maximal cutoff $\Lambda=400$~MeV, both binding energies account for roughly 80\% of the respective experimental values. Including the three-body force improves the description of both nuclei. While the $^{3}$H energy $E_{\rm 2NF+3NF}^{\prime}(^{3}$H) trivially reproduces the experimental value by construction, the predicted $^{4}$He energy $E_{\rm 2NF+3NF}^{\prime}(^{4}$He) exhibits slight overbinding for all $\Lambda$ values. This deviation from experiment gradually decreases as $\Lambda$ increases, asymptotically approaching the experimental value for large $\Lambda$.

The third part of Table~\ref{tab:LECs} presents results calculated with the GIR terms fully taken into account.
We can estimate the quantitative impact of the GIR terms by comparing predictions from the 2NF only. Note that the energies $E_{\rm 2NF}^{\prime}$ and $E_{\rm 2NF}$ are calculated using the same LECs $B_{1-2}$ and $C_{1-7}$ calibrated in the center-of-mass frame. However, $E_{\rm 2NF}$ contains additional corrections from the GIR terms. Therefore, we directly evaluate the GIR contribution as the difference $E_{\rm GIR} = E_{\rm 2NF} - E_{\rm 2NF}^{\prime}$. For the maximal cutoff $\Lambda=400$~MeV, this GIR correction amounts to $0.2$~MeV for $^{3}$H and $1.0$~MeV for $^{4}$He, which is highly suppressed. We expect that for even larger cutoffs (\textit{e.g.}, $\Lambda=450$, $500$~MeV) the GIR corrections would be further reduced according to $\mathcal{O}(\Lambda^{-2})$.
Conversely, for lower $\Lambda$-values, equivalent to larger lattice spacings, we observe significant Galilean invariance breaking effects. 
This is reflected by larger GIR contributions to the binding energies. 
For the minimal cutoff $\Lambda=250$~MeV, the GIR corrections contribute $1.2$~MeV for $^{3}$H and $5$~MeV for $^{4}$He, approximately 5--6 times larger than for $\Lambda=400$~MeV. Over the $\Lambda$ interval considered, the magnitude of these corrections is consistent with the naive expectation that the GIR operators belong to the $Q^{2}$-order (NLO). For even lower cutoffs, we anticipate more severe Galilean invariance breaking effects, where the required corrections might invalidate the naive power-counting.

The results labeled $E_{\rm 2NF+3NF}$ include contributions from both the GIR terms and the three-body force. Compared to the results without GIR corrections (primed values), the three-body LEC $c_{E}$ is readjusted for each $\Lambda$ to ensure exact reproduction of the $^{3}$H binding energy. This readjustment completely cancels the effect of the GIR terms in $^{3}$H due to the fitting strategy. However, the interplay between these terms leaves non-negligible net effects on the predicted $^{4}$He energies. Remarkably, $E_{\rm 2NF+3NF}(^{4}$He) is almost constant against varying $\Lambda$ and precisely reproduces the experimental value within statistical errors of order 100 keV. Comparing the results with and without GIR corrections ($E_{\rm 2NF+3NF}$ v.s. $E_{\rm 2NF+3NF}^\prime$)  demonstrates a significant improvement in RG-invariance.

\begin{figure}[htbp]
\centering
\includegraphics[width=0.45\textwidth]{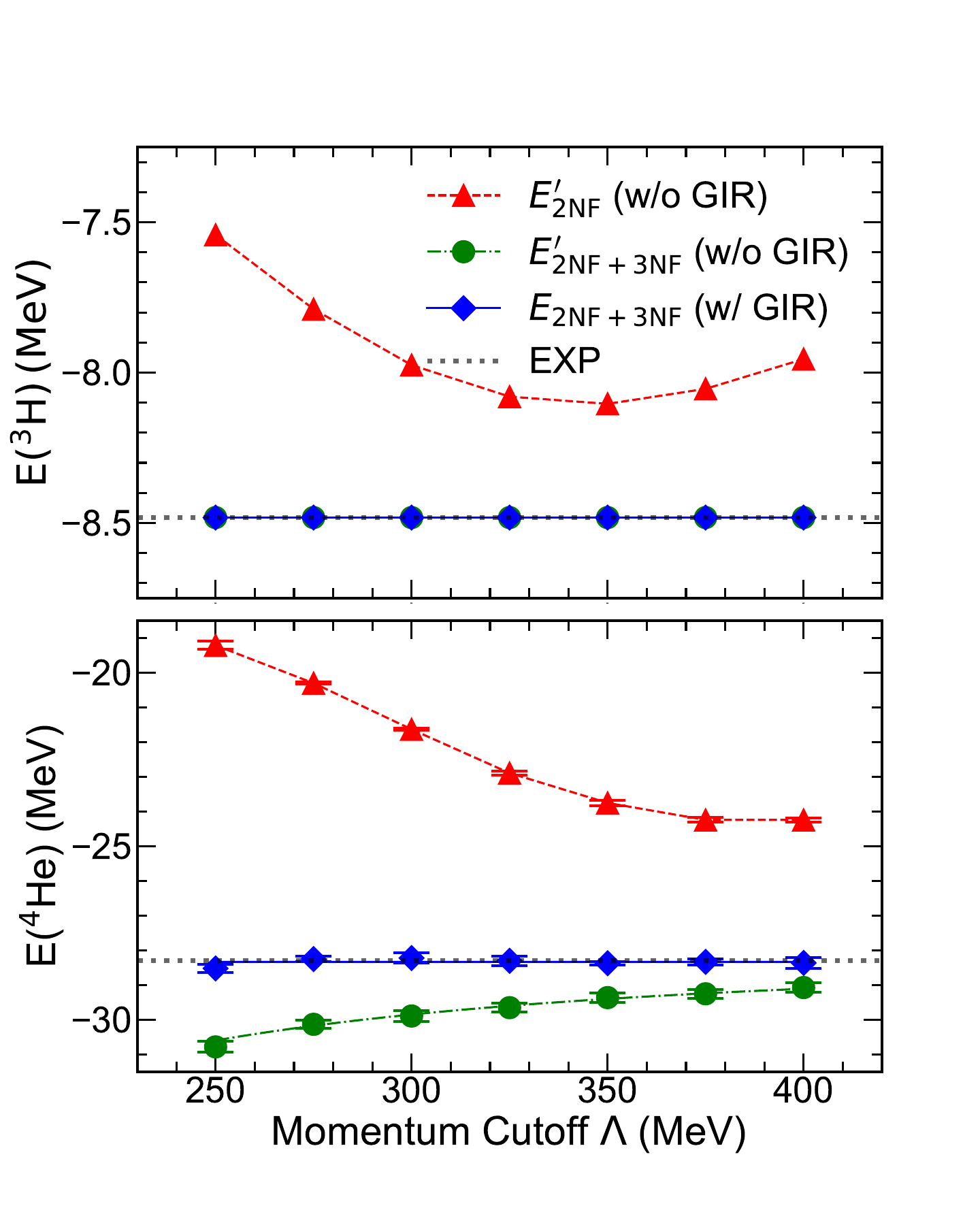}
\captionsetup{justification=raggedright, singlelinecheck=false}
\caption{\label{fig:fig1} Binding energies of $^3$H (Upper panel) and $^4$He (Lower panel) calculated with two-body force only (triangles) and two- plus three-body forces (circles/diamonds) as functions of $\Lambda$. Diamonds represents the results including the GIR terms.
Dotted lines denote the experimental values.}
\end{figure}

The trends observed in the numerical results are more clearly visualized in Fig.~\ref{fig:fig1}, which displays the $^{3}$H (upper panel) and $^{4}$He (lower panel) binding energies versus $\Lambda$. Triangles, circles, and diamonds denote the quantities $E_{\rm 2NF}^{\prime}$, $E_{\rm 2NF+3NF}^{\prime}$, and $E_{\rm 2NF+3NF}$ from Table~\ref{tab:LECs}, respectively. For $^{3}$H, the results $E_{\rm 2NF}(^{3}$H) calculated without the three-body force vary significantly with $\Lambda$, indicating a severe violation of RG-invariance. The results $E_{\rm 2NF+3NF}^{\prime}$ and $E_{\rm 2NF+3NF}$, which include the three-body force, coincide with the experimental value, consistent with the fitting strategy. For $^{4}$He, the results are pure predictions. The result $E_{\rm 2NF}^{\prime}(^{4}$He) shows a strong $\Lambda$-dependence similar to $^{3}$H. Including the three-body force yields $E_{\rm 2NF+3NF}^{\prime}(^{4}$He), which is a smooth function of $\Lambda$ converging towards the experimental value. The results $E_{\rm 2NF+3NF}(^{4}$He) including both GIR terms and the three-body force show excellent agreement with experiment for all values of $\Lambda$.

\begin{figure}[htbp]
\centering
\includegraphics[width=0.45\textwidth]{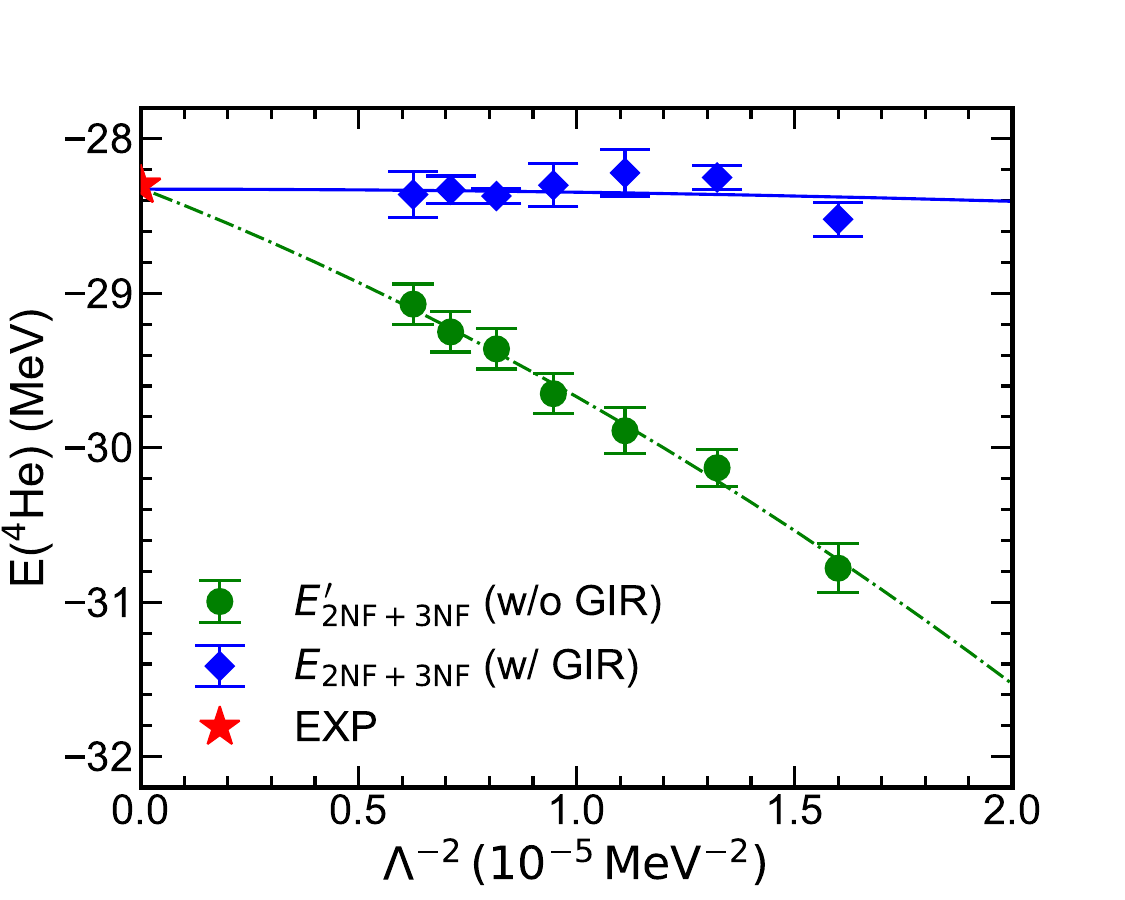}
\captionsetup{justification=raggedright, singlelinecheck=false}
\caption{\label{fig:fig--2505-1} Calculated ground-state energies of $^4$He as functions of $\Lambda^{-2}$.
Circles (diamonds) denote results calculated without (with) the GIR terms.
Lines represent the extrapolations according to Eq.~(\ref{eq:asymptotic}). Red star marks the experimental value.}
\end{figure}

We further examine the asymptotic behavior at large $\Lambda$. Fig.~\ref{fig:fig--2505-1} plots the predicted $^{4}$He energies with ($E_{\rm 2NF+3NF}$) and without ($E_{\rm 2NF+3NF}^{\prime}$) the GIR terms as functions of $\Lambda^{-2}$. The lines represent the function
\begin{equation}
   E(\Lambda) = E(\infty) + \frac{c_2}{\Lambda^2} + \frac{c_4}{\Lambda^4}, \label{eq:asymptotic}
\end{equation}
where $c_2$ and $c_4$ are parameters fitted to the Monte Carlo results. The $c_2$- and $c_4$-terms originate from $Q^{2}$-order and $Q^{4}$-order contact operators generated by the sliding cutoff, respectively. In Fig.~\ref{fig:fig--2505-1}, both groups of results show rather weak dependencies on the $c_4$-term. While the results without GIR (dots) exhibit an approximately linear dependence on $\Lambda^{-2}$, signifying a pronounced $c_2$-term, the GIR-corrected results (diamonds) are consistent with $c_2 = c_4 = 0$. Significantly, the GIR terms completely cancel the $c_2$-term in Eq.~(\ref{eq:asymptotic}), leaving a cutoff-independent prediction. As the $\Lambda$-dependent terms in Eq.~(\ref{eq:asymptotic}) diminish asymptotically, we extrapolate to $\Lambda\rightarrow\infty$ for each group. We obtain $E(\infty) = -28.32(33)$~MeV without GIR and $E(\infty) = -28.33(6)$~MeV with GIR, both coinciding with the experimental value $E_{\rm exp} = -28.3$~MeV.

\begin{figure}[ht]
\centering
\includegraphics[width=0.45\textwidth]{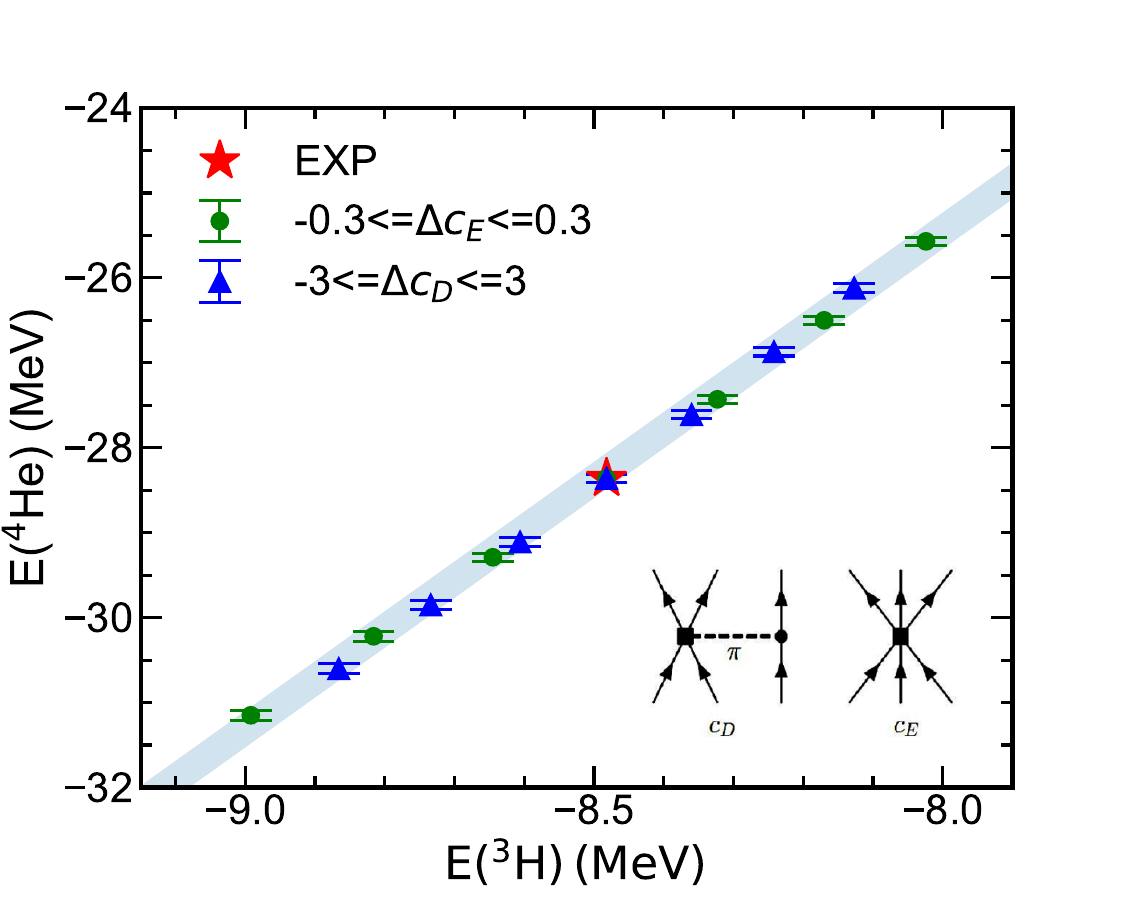}
\captionsetup{justification=raggedright, singlelinecheck=false}
\caption{\label{fig:cDcE} Binding energies of $^3$H and $^4$He calculated with varying three-body force LECs $c_D$ and $c_E$. 
The insets shows the leading 3NF diagrams with adjustable parameters $c_D$ and $c_E$.
The calculations cover the ranges $\triangle c_\mathrm{E} \in$ [$-$0.3,0.3] and $\triangle c_\mathrm{D} \in$ [$-$3,3].
Red star marks the experimental values.
Empirical Tjon band is depicted to guide the eyes.}
\end{figure}

Finally, we investigate whether the observed pattern holds with chiral 3NFs.
In the chiral expansion, two independent adjustable 3NF parameters exist at N$^{2}$LO. 
Conventionally, the term proportional to $c_E$ takes the form in Eq.~(3), while the term proportional to $c_D$ describes the leading-order one-pion-exchange 3NF~\cite{van1994few, PRC59-53, PhysRevC.66.064001, EPJA41-125}. 
Previous calculations with chiral EFT regulated in the continuum found these two terms highly correlated for light-nuclei observables~\cite{PLB56-217, PRC20-340, PLB607-254}.
Consequently, accurately determining their values using few-body data alone is difficult, thus modern \textit{ab initio} calculations often employ medium-mass nuclei or nuclear matter as extra constraints~\cite{PRC100-024318, PRL122-042501, PLB808-135651, PRC102-034313}.
We start from a calculation at $\Lambda=350$~MeV with both GIR and the $c_E$-term included ($c_D=0$). In this case, $^{3}$H and $^{4}$He energies are nicely reproduced. Fig.~\ref{fig:cDcE} shows the correlation of binding energies calculated by varying $c_E$ and $c_D$ independently.
Here, $\Delta c_E$ and $\Delta c_D$ denote deviations from the central values $c_E=0.245$ and $c_D=0$ in Table~\ref{tab:LECs}.
The results form a Tjon band~\cite{PLB56-217} passing through the experimental value, implying that combinations of 3NFs reproducing $E(^{3}$H) also reproduce $E(^{4}$He).
Thus, we conclude that the RG-invariance improvement observed with a simple contact 3NF robustly persists for more general 3NFs.

Having included charge-symmetry breaking effects, we also calculate $^{3}$He binding energies with both GIR and 3NFs. Table~\ref{tab:LECs} shows these results as $E_{\rm 2NF+3NF}(^{3}$He) and the energy splitting $\Delta(^{3}$H-$^{3}$He). These observables are well reproduced, with discrepancies relative to experiments of only $\sim$10~keV. We also calculated charge radii for these light nuclei, observing a convergence pattern towards the experimental values. Achieving an RG-invariant prediction for radii would require supplementing the density operators with cutoff-dependent corrections, which is beyond the scope of this work.

$\phantom{}$



\section{Conclusions and outlook}

Recent years have witnessed significant progress in nuclear lattice effective field theory (NLEFT) as a successful first-principles method for nuclear structure. 
Improvements in many-body algorithms and lattice chiral forces now enable many high-precision \textit{ab initio} calculations, ranging from the Hoyle state to nuclear thermodynamics. 
Faithfully reproducing experimental results and reliably extrapolating to unknown regions remain central challenges at the frontiers of the field. 
On the other hand, the high-precision techniques also allow careful examination of result dependencies on unphysical artifacts, which is essential to quantify theoretical uncertainties. 
In NLEFT, the most significant artifacts arise from lattice discretizations. 
Quantifying and systematically investigating the dependency of calculations on lattice cutoff is therefore a central problem requiring specific attentions.

We performed high-precision calculations for few-body observables ($A \leq 4$) employing state-of-the-art lattice algorithms. The results directly verify predictions derived from the Wilsonian renormalization group (RG) based on an N$^2$LO chiral interaction. Crucially, as the lattice regulator breaks symmetries such as Galilean invariance, varying the cutoff (inverse lattice spacing) induces additional contact terms that explicitly violate these symmetries. Therefore, a fully RG-invariant calculation requires incorporating these terms into the Hamiltonian. Using the $^{4}$He binding energy as an example, we demonstrate that Galilean invariance restoration terms are essential both for preserving RG invariance and for reproducing experimental results. Significantly, these RG-invariant results coincide with experiment, establishing NLEFT's capability for reliable, parameter-free predictions. 
While results shown here are limited to soft cutoffs $\Lambda \leq 400$~MeV, calculations extending to about $500$~MeV are straightforward and are not expected to yield qualitative differences.
Notably, calculations omitting symmetry restoration exhibit asymptotic behavior with slow convergence, thus the RG-invariant result can also be extracted by extrapolating to the continuum limit ($\Lambda \to \infty$).

In this work, we employed a soft regulator focusing exclusively on Galilean invariance breaking. 
Our findings are directly extendable to standard lattice regulators as used in Ref.~\cite{klein2018tjon}, where rotational symmetry breaking effects also contribute.
Rotational invariance restoration terms can be introduced analogously, with their corresponding low-energy constants (LECs) determined using the condition of spatial isotropy.
We anticipate that the Hamiltonian incorporating the full restoration of these fundamental symmetries will exhibit significantly improved RG invariance, a prediction requiring high-precision numerical verification. Such verification poses considerable challenges, particularly for many-body systems. Research addressing these challenges is actively underway.

\section*{Acknowledgements}
We thank members of the Nuclear Lattice Effective Field Theory Collaboration for insightful discussions.
This work has been supported by NSAF No. U2330401 and National Natural Science Foundation of China with Grant Nos. 12275259, 12405143, 12547105.

\appendix



\bibliographystyle{elsarticle-harv} 






\end{document}